\documentclass[]{article}

\usepackage[sc]{mathpazo} 
\usepackage[T1]{fontenc} 
\usepackage{microtype} 

\usepackage[english]{babel} 

\usepackage[hmarginratio=1:1,top=32mm,columnsep=20pt]{geometry} 
\usepackage[hang, small,labelfont=bf,up,textfont=it,up]{caption} 
\usepackage{booktabs} 

\usepackage{lettrine} 

\usepackage{enumitem} 
\setlist[itemize]{noitemsep} 

\usepackage{abstract} 

\usepackage{titlesec} 
\titleformat{\section}[block]{\large\scshape\centering}{\thesection.}{1em}{} 
\titleformat{\subsection}[block]{\large}{\thesubsection.}{1em}{} 

\usepackage{fancyhdr} 
\usepackage{titling} 

\usepackage{hyperref} 

\usepackage{epsfig}
\usepackage{graphicx}	
\usepackage{amsmath}	
\usepackage{amssymb}	
\usepackage{float}
\usepackage{subcaption}

\pretitle{\begin{center}\Large\bfseries} 
	\posttitle{\end{center}} 
\title{Dark energy and the fitting problem} 
\author{%
	Vincent Deledicque \\[1ex] 
	\normalsize \textit{No affiliation} \\ 
	\normalsize \href{mailto:vincent.deledicq@gmail.com}{vincent.deledicq@gmail.com} 
}
\date{\today} 
\renewcommand{\maketitlehookd}{%
	\begin{abstract}
		\noindent At a global scale, the universe is generally fitted by an idealized manifold described by the FLRW metric. This is in particular the case when probing the universe to determine its dynamics. The process that fits the idealized manifold to the real universe is however not uniquely defined. This process may depend on the cosmic probe that has been used for the measurements, and could hence lead to different observed temporal evolutions of the scale factor. A correct interpretation of the observed accelerated expansion of the universe requires therefore first a thorough understanding of the fitting process that has been implicitly applied. In this article we establish the fitting processes for the SNIa, BAO and CMB cosmic probes, and deduce the related averaged Einstein equations. We demonstrate that the way these fittings have been applied in practice lead to an apparent dark energy effect. We also highlight the conceptual differences in the fitting processes between the SNIa and BAO probes on the one hand, and the CMB probe on the other hand. Considering those differences, we then show how the so-called Hubble tension can be explained.
	\end{abstract}

}

\begin{document}
	
	\maketitle
	
	
	\section{Introduction}	
	Most of the time, real manifolds present complicated features. Fortunately, at a global scale, they can often be approximated by an idealized manifold, described by a limited number of parameters. But in general, the approximation that can be performed is not uniquely defined. The fitting problem, consisting in determining how an idealized manifold should be best fitted onto the real one, is fundamental. Indeed, being considered to be representative of the real manifold, the fitted idealized manifold is the one that will be used in practice, since its description is considerably simplified. So, any property deduced from the fitted manifold will be assumed to be applicable in the same way on the real manifold. It is therefore important, when analysing some given real manifold, to establish the most adequate fitting process for its idealized representation, ensuring that this representativeness is meaningful.
	
	To understand the underlying implications of the fitting problem, let us start our examination with a simple illustrative example, i.e., the case of the earth. It is known that the earth is not a perfect sphere. Mountains and abysses form small perturbations, and the earth is also slightly flattened at the poles. However, since globally it looks like a sphere, it makes sense to approximate it as such, because its whole geometry would then be described by one single parameter, i.e., a mean radius. This considerably simplifies its description. But several fitting processes can be used to approximate the earth as a sphere, each one leading possibly to a different mean radius. The earth's mean radius is hence purely conventional, and depends on the way it has been defined (and measured). For example, the International Union of Geodesy and Geophysics (IUGG) provides three reference values based on three different definitions, see \cite{Moritz}. The first one corresponds to the average of three specific radii (two on the equator and one at a pole), the second one corresponds to the radius of a sphere with the same surface area, and the last one corresponds to the radius of a sphere having the same volume. These three definitions thus imply different measurements, leading to different fittings of the idealized sphere. Other definitions could be envisaged. Fortunately, in the case of the earth, differences between various definitions are quite insignificant. This is because the perturbations with respect to what would be a perfect spherical earth are of the order of 10 km, to be compared with a mean radius of about 6370 km. The earth is a trivial example, but it will be helpful to regularly refer to it in the following to better understand more complex situations.
	
	In a similar way, the universe corresponds to a four-dimensional manifold, looking globally like an idealized manifold described by the Friedmann-Lema\^\i tre-Robertson-Walker (FLRW) metric. Here also, local perturbations exist and imply that this FLRW metric can only be considered as an approximation of the real one. Obviously, for practical reasons, this approximation makes sense, since it allows to study the dynamics of the universe by using one single evolving parameter, namely the scale factor $a$. Unfortunately, with respect to the earth, the universe presents several major difficulties. Firstly, as said, the scale factor is not a constant, but is expected to evolve over time. Secondly, a curvature parameter $k$ needs also to be taken into account. Thirdly, perturbations of the real manifold with respect to the FLRW manifold could be more important than in the case of the earth (see further), and this could give room for a wider variety of fitting processes (and hence different dynamics). Fourthly, being four-dimensional, it is much more difficult to represent, even in a schematic way, the related manifold. In particular, it would not be self-evident to detect that some approximate manifold has been fitted inadequately. Lastly, whereas there is no fundamental physical law that allows us to predict the earth radius, the as measured evolution of the scale factor can be compared with a theoretical prediction, and this leads to additional difficulties.
	
	Let us elaborate a little more on this last difficulty. Theoretically, the dynamics of the universe is expected to be related in some way to the Einstein equation of General Relativity. However, determining this relation could be more difficult than generally thought. When establishing the Friedmann equation, we assume that the Einstein equation can be applied as such at the cosmic scale, but this is a very strong assumption. We should indeed remind that the Einstein equation has been proven to be valid at what we could call a local scale, but it has not been proven to be valid at the cosmic scale. In fact, theoretically, being a field equation, each tensor of the Einstein equation should be evaluated at each specific point. When considering that tensors and equations can be evaluated at a macroscopic scale, we perform some averaging process. But as is well known, averaging field equations can lead to extra terms (as an example, averaging the Navier-Stokes in fluid mechanics leads to additional terms accounting for the turbulence effects). Unfortunately, it is still unclear how the Einstein equation should be averaged, see for example \cite{Hoogen}. But perhaps more importantly, we should stress that the way the Einstein equation should be averaged should in fact depend on the fitting process itself. Obviously, different fitting processes will lead to different idealized manifolds, related to different dynamics of their FLRW metric. We should thus indeed expect that they are described by different averaged Einstein equations.
		
	Acknowledging that different fittings could lead to different dynamics, we understand the importance to address the fitting problem with the required attention in the case of the universe. The importance to address this problem has already been highlighted by other authors, see for example \cite{Clifton} or more recently \cite{Mohayaee}. To be complete, the fitting problem should be addressed in two steps. The first step consists in understanding how the FLRW metric can be (or has been) fitted onto the real universe. This fitting process is directly related to the probe technique that has been used. Reminding indeed the case of the earth mentioned above, the first probe, based on a direct measure of three radii, leads to a different mean radius than the two other probes, which are based on indirect measurements. In a similar way, we should expect that the probe techniques used in cosmology, in particular the SNIa probe, the Baryon Acoustic Oscillations (BAO) probe, and the Cosmic Microwave Background (CMB) probe, could lead to different fittings of the FLRW metric onto the real universe. So, each probe technique has its own fitting process, and should thus be examined separately. Then, knowing the fitting process for a specific probe technique, in a second step, the Einstein equation of General Relativity should be adequately averaged to predict the expected dynamics of the as fitted manifold. 
	
	Studies performed up to now were generally either focused on the first step only (what would be the best fitting process, see in particular \cite{Ellis}), either on the second step only (what would be the best averaging process of the Einstein equation, see for example \cite{Hoogen}, \cite{Wiltshire} or \cite{Zalaletdinov}). As an illustration, the study of backreaction effects, which has received considerable attention by some research teams, is part of this second step. According to this theory, the Einstein equation of General Relativity is not applicable as such on a global scale, and needs to be averaged somehow. The proposed averaging process is non-linear, which leads to the appearance of a new term in the averaged Einstein equation, accounting according to that theory for the dark energy effect. However, the study of backreaction effects does not relate the proposed averaging process to a specific fitting process. So it is implicitly assumed that the various probes of the universes are related to an identical fitting process, and that this one is characterized by the proposed average Einstein equation. But the link between the proposed averaging process and the different probes has not been demonstrated. In other words, it has not been shown that the SNIa, BAO, CMB and all other possible probes are indeed characterized by such an averaging process. Moreover, the need to investigate the fitting problem is in some way also supported by the fact that some cosmic probes lead to small but still significant differences in the observations (this is the so-called Hubble tension). The root cause of this Hubble tension is currently a matter of research, but reminding once again the case of the earth, the Hubble tension could simply be explained by the existence of different fitting processes related to those cosmic probes. We will justify that this is indeed the case.
	
	To summarize, in their interpretation, results from the different cosmic probes have been taken as such, without really wondering on which approximate manifold of the real universe we were working, and if this one could effectively be considered as globally representative of the real universe. But it is essential to understand which fitting process characterizes each probe. And this is what we propose to investigate in this article. Therefore, we need to understand that the two steps mentioned above are closely related, and should be analysed together. A specific averaging process of the Einstein equation of General relativity is only applicable to some specific fitting process, and vice versa. 
	
	In section $\ref{Ao}$ we introduce the two major averaging operations that need to be carried out when analysing the results of cosmic probes. The first one consists in averaging adequately measurement results (corresponding to what we observe), and the second one consists in averaging the Einstein equation of General Relativity (corresponding to what we predict). For this last one, we develop a simple averaging operation based on the requirement that it be covariant and that it allows to link the idealized and the real FLRW manifolds. This averaged Einstein equation is indeed expected to be not uniquely defined, and should contain a term that depends on the fitting process. In section $\ref{S1}$ we then investigate the SNIa and BAO cosmic probes. It will be justified that these probes are indeed related to an identical fitting process. We will determine this fitting process, and use it to deduce the related averaged Einstein equation. We will demonstrate that this equation involves a term that accounts for an apparent dark energy effect, able to explain the observed accelerated expansion of the universe, as evidenced by \cite{Riess} and \cite{Perlmutter}. In section $\ref{S2}$ we finally investigate the CMB probe. Similarly to the previous section, we will examine how this probe implicitly defines its fitting process, and then show how it affects the average Einstein equation. We will also explain why the way we apply the CMB probe leads to the Hubble tension with the two previous probes.
	

	\section{Averaging operations}\label{Ao}
	
	In general, fitting processes imply averaging operations. This is because each local property could be affected by a perturbation with respect to the average value, and could hence generally not be considered as representative of the global behaviour. Therefore, measurements are carried out at multiple locations so that existing perturbations are averaged over, in order to capture only the mean value of the parameters defining the idealized manifold. In the case of the universe, we do not escape to this common practice. Moreover, since we expect that the scale factor evolves over time, probe techniques generally (but not always) measure an average value at different times. This is the case for example for measurements performed on SNIa. Averaging the results obtained by some probe technique constitutes the first averaging operation that has to be performed when investigating the dynamics of the universe. This first averaging operation is thus related to the measurement.
	
	It is important to understand that each specific probe technique implicitly defines its own fitting process. Reminding the case of the earth, measuring directly the mean radius, or measuring first an mean surface or volume and then deducing the mean radius leads to different results. This thus also means that each specific probe technique will be related to a particular averaging operation in the measurements. In fact, the fitting process can be directly characterized by the way this first averaging operation has been performed, and this will be an extremely useful information. To better understand what this means in practice, it will be helpful to start with the second averaging operation that has to be performed when investigating the dynamics of the universe, i.e., the averaging of the Einstein equation of General Relativity. This second averaging operation is thus related to the theoretical prediction.
	
	As said in the introduction, the as measured evolution of the scale factor is expected to be related somehow to the Einstein equation of General Relativity. But this last equation is valid locally, and applying it to the universe requires some averaging operation on it. When averaging the Einstein equation, we need to be careful, because such operation implies an integration over some sub-manifold. It is indeed well-known that an integration (i.e., a summation) over curved manifolds is in general well-defined for scalars only. On the contrary, integrating tensors is a questionable operation. Adding a tensor located at some point to a tensor located at another point requires to parallel transport one of them to the other one along some curve. But in general, the tensor obtained after being parallel transported is dependent upon the selected curve, meaning that the result of the summation is not well defined. We could thus fear conceptual issues when integrating the tensors appearing in the Einstein equation when trying to average it. To avoid such issues, the averaging process we propose here is performed by integrating scalars only. In fact, there is no need to obtain an averaged Einstein equation in tensorial form. If the curvature parameter $k$ is known, the FLRW metric is defined only by the scale factor, therefore, one single average scalar equation is sufficient to predict its dynamics. This also renders the equation automatically covariant. We finally also require the averaged Einstein equation to take into account the characteristic of the averaging operation of the considered fitting process. Indeed, since the averaged Einstein equation should depend on the fitting process, we expect that it contains some term making the link between the real and the idealized manifolds, which, as we will see, can be defined by the averaging operation on the measurements.
	
	This averaging process for the Einstein equation is established as follows. Let us first define the notations that we will use: $g_{\mu\nu}$ is the local metric tensor, and $\overline{g}_{\mu\nu}$ is the FLRW metric tensor obtained from some specific fitting process. We then define the local perturbation in the metric tensor $\Delta g_{\mu\nu}$ such that
	\begin{equation}
		g_{\mu\nu} = \overline{g}_{\mu\nu} + \Delta g_{\mu\nu}\,.
	\end{equation}
	We also define the inverse FLRW metric tensor $\overline{g}^{\mu\nu}$ such that
	\begin{equation}
		\overline{g}^{\alpha\mu}\overline{g}_{\mu\beta} = \delta^\alpha_{\ \beta}\,.
	\end{equation}
	In the same way, $G_{\mu\nu}$ is the local Einstein tensor, and $\overline{G}_{\mu\nu}$ is the FLRW related Einstein tensor. The difference between those two last tensors is written as $\Delta G_{\mu\nu}$:
	\begin{equation}
		G_{\mu\nu} = \overline{G}_{\mu\nu} + \Delta G_{\mu\nu}\,.
	\end{equation}
	
	In the following, we will make use of a specific frame of reference based on the co-moving coordinates, for which $t$ is the cosmological time coordinate, and where $(x,y,z)$ are the spatial Cartesian coordinates. In these specific coordinates, $\overline{g}_{\mu\nu}$ is diagonal. Assuming $k=0$, we have $\overline{g}_{tt} = -1$, $\overline{g}_{ii} = a^2$ where $i = x,y,z$, and all other components are zero. Similarly, $\overline{g}^{tt} = -1$, $\overline{g}^{ii} = 1/a^2$ and all other components of $\overline{g}^{\mu\nu}$ are zero. Finally, $\overline{G}_{tt} = 3\dot{a}^2/a^2$, $\overline{G}_{ii} = -2a\ddot{a}-\dot{a}^2$, and all other components of $\overline{G}_{\mu\nu}$ are zero. 
	
	Assuming that the cosmological constant $\Lambda$ is zero, the Einstein equation of General Relativity reads
	\begin{equation} \label{Einsteinequation}
		G_{\mu\nu} = 8\pi G T_{\mu\nu}\,.
	\end{equation}	
	Now, given the FLRW metric and the tensors derived from it, we note that together with the Einstein equation, we can produce two independent scalar relations. A first possibility consists in multiplying Eq.\ $(\ref{Einsteinequation})$ by $\overline{g}^{\mu\nu}$:
	\begin{equation}
		G_{\mu\nu}\overline{g}^{\mu\nu} = 8\pi G T_{\mu\nu}\overline{g}^{\mu\nu}\,.
	\end{equation}
	This indeed constitutes a scalar relation. We may hence integrate it over space (meaning over all events that occur at the same cosmological time):
	\begin{equation}\label{dget}
		\int_V G_{\mu\nu}\overline{g}^{\mu\nu}\sqrt{|g_{ij}|}dV = \int_V 8\pi G T_{\mu\nu}\overline{g}^{\mu\nu}\sqrt{|g_{ij}|}dV\,,
	\end{equation}
	where $|g_{ij}|$ is the magnitude of the determinant of the spatial part of $g_{\mu\nu}$, and where $\int_V$ means an integration over a spatial volume $V$ supposed to be sufficiently large so that it can be considered as representative of whole space.
	
	A second possibility to obtain a scalar relation from the Einstein equation of General Relativity and from the FLRW metric consists in multiplying Eq.\ $(\ref{Einsteinequation})$ by $\overline{g}^{\mu\alpha}\overline{g}^{\nu\beta}\overline{G}_{\alpha\beta}$. Integrating this scalar relation over space leads to:
	\begin{equation}\label{dget2}
		\int_V G_{\mu\nu}\overline{g}^{\mu\alpha}\overline{g}^{\nu\beta}\overline{G}_{\alpha\beta}\sqrt{|g_{ij}|}dV = \int_V 8\pi G T_{\mu\nu}\overline{g}^{\mu\alpha}\overline{g}^{\nu\beta}\overline{G}_{\alpha\beta}\sqrt{|g_{ij}|}dV\,,
	\end{equation} 

	Since the equations $(\ref{dget})$ and $(\ref{dget2})$ are scalar relations, they are frame invariant, and we may evaluate them in a specific frame of reference without worry. We will therefore use the specific frame of reference mentioned above. Assuming that space is globally isotropic, we admit that on average the different spatial diagonal components of $G_{\mu\nu}$ are identical, and will be written as $G_{xx}$. Developing Eq.\ $(\ref{dget})$, we then get
	\begin{equation}\label{dget3}
		\int_V\left(\frac{3}{a^2}G_{xx}-G_{tt}\right)\sqrt{|g_{ij}|}dV = \int_V 8\pi G \left(\frac{3}{a^2}T_{xx} - T_{tt}\right)\sqrt{|g_{ij}|}dV\,.
	\end{equation}
	In the same way, developing Eq.\ $(\ref{dget2})$ gives
	\begin{eqnarray}\label{dget4}
		&&\int_V\left[3\left(-2\frac{\ddot{a}}{a^3} - \frac{\dot{a}^2}{a^4}\right)G_{xx} + 3\frac{\dot{a}^2}{a^2}G_{tt}\right]\sqrt{|g_{ij}|}dV \nonumber
		\\
		&& = \int_V 8\pi G \left[3\left(-2\frac{\ddot{a}}{a^3} - \frac{\dot{a}^2}{a^4}\right)T_{xx} +3\frac{\dot{a}^2}{a^2}T_{tt} \right]\sqrt{|g_{ij}|}dV\,.
	\end{eqnarray}
	Since the scale factor and its temporal derivatives do not depend on the spatial coordinates, they can be considered as a constant in the integrations. We then multiply Eq.\ $(\ref{dget3})$ by $3\dot{a}^2/a^2$ and add Eq.\ $(\ref{dget4})$ to it:
	\begin{equation}
		\int_V6\left(\frac{\dot{a}^2}{a^4} - \frac{\ddot{a}}{a^3}\right)G_{xx}\sqrt{|g_{ij}|}dV = \int_V 6\left(\frac{\dot{a}^2}{a^4} - \frac{\ddot{a}}{a^3}\right)8\pi G T_{xx}\sqrt{|g_{ij}|}dV\,,
	\end{equation}
	which can be simplified:
	\begin{equation}\label{tpt1x}
		\int_V G_{xx}\sqrt{|g_{ij}|}dV = 8\pi G \int_V T_{xx}\sqrt{|g_{ij}|}dV\,.
	\end{equation}
	Using then this latter relation together with Eq.\ $(\ref{dget3})$, we deduce that
	\begin{equation}\label{tpt2x}
		\int_V G_{tt}\sqrt{|g_{ij}|}dV = 8\pi G \int_V T_{tt}\sqrt{|g_{ij}|}dV\,.
	\end{equation}

	It is important to stress that if Eq.\ $(\ref{tpt1x})$ and $(\ref{tpt2x})$ could give the impression that we integrate tensors over space, this is in fact not the case: they are truly scalar relations, as can be seen from the way we derived them. Their scalar nature is hidden by the simplifications we made. So, these relations are well defined.
	
	Writing finally $G_{\mu\nu}$ in function of its fitted part and its perturbation part, we get
	\begin{equation}\label{tpt1}
		\int_V\left(\overline{G}_{xx} + \Delta G_{xx}\right)\sqrt{|g_{ij}|}dV = 8\pi G \int_V T_{xx}\sqrt{|g_{ij}|}dV
	\end{equation}
	and
	\begin{equation}\label{tpt2}
		\int_V\left(\overline{G}_{tt}+\Delta G_{tt}\right)\sqrt{|g_{ij}|}dV = 8\pi G \int_V T_{tt}\sqrt{|g_{ij}|}dV\,.
	\end{equation}

	Let us in particular examine Eq.\ $(\ref{tpt2})$. Since $\overline{G}_{tt}$ does not depend on the spatial coordinates, it can be taken out of the integral, and this last equation can be written as
	\begin{equation}\label{az}
		\overline{G}_{tt} + \frac{\int_V\Delta G_{tt}\sqrt{|g_{ij}|}dV}{\int_V \sqrt{|g_{ij}|}dV} = 8\pi G\frac{\int_V  T_{tt}\sqrt{|g_{ij}|}dV}{\int_V \sqrt{|g_{ij}|}dV}\,.
	\end{equation}
	In the right hand side, we recognize the average density over space:
	\begin{equation}\label{ro}
		\rho = \frac{\int_V  T_{tt}\sqrt{|g_{ij}|}dV}{\int_V \sqrt{|g_{ij}|}dV}\,.
	\end{equation}
	So Eq.\ $(\ref{az})$ can be written as
	\begin{equation}\label{qss}
		3\frac{\dot{a}^2}{a^2} + \frac{\int_V\Delta G_{tt}\sqrt{|g_{ij}|}dV}{\int_V \sqrt{|g_{ij}|}dV} = 8\pi G\rho\,.
	\end{equation}
	This looks like the Friedmann equation, except that an additional term appears in the left hand side. This term is directly related to the spatial average of the perturbations in the Einstein tensor, and depends on the way the FLRW metric has been fitted. Indeed, the fitting process will directly determine if the perturbations $\Delta G_{tt}$ are on average mostly positive, negative or even cancel. If the fitting process is such that these perturbations cancel on average, then the average Einstein equation reduces to the Friedmann equation. If not, this additional term will be responsible for a global dynamics of the fitted manifold that deviates from the Friedman equation. This additional term is the one we expected to make the link between the idealized and real manifolds, and so for each specific probe, we need to determine its value.
	
	Let us make here the following comment. The additional term is defined on the basis of a spatial integral. To determine it quantitatively, we need to obtain data which are spread over whole space. In practice, the events for which we have measurement results and which occur at the same cosmological time are not spread over the whole three-dimensional space, but are instead generally located on an almost spherical two-dimensional manifold centred on the earth. Therefore, in practice, the averaging operation of the Einstein equation should best be carried out on this sphere instead over a spatial volume, as we did above. Fortunately, since space is assumed to be globally homogeneous and isotropic, these measurements are supposed to be representative for whole space at the considered cosmological time. Therefore, the averaging operation is expected to be identical if we perform it on a volume or on a surface. So Eq.\ $(\ref{qss})$ is considered to be perfectly valid when used as comparison with the dynamics deduced form the measurements.
	
	To synthesize in a few words what we have done here: knowing the local dynamics (described by the Einstein equation of General Relativity), and taking into account how the fitted and real manifold are correlated on average through their respective Einstein tensors, we have deduced an equation able to predict the dynamics of the fitted manifold. This equation is named to be the average Einstein equation, by using the usual terminology, but would perhaps better be called the fitted Einstein equation, to clearly highlight that different fitting processes will have different fitted Einstein equations. Though, we will continue to use the usual terminology here.
	
	At this stage, it is instructive to have an idea about the smoothness of the real metric (or more particularly of the Einstein tensor, because this is the one that describes the dynamics of the universe), and this can be examined by evaluating the significance of the perturbations with respect to an idealized FLRW metric and its related Einstein tensor. This will allow us to know how far different fitting processes could deviate from each other. Obviously, locally (especially in the vicinity of massive objects), we can have extremely important deviations of the real universe with respect to the idealized average one, but those perturbations are very localized and should not impact the averaging operation significantly, which implies integrations over large spaces. So we are not interested here in such very local perturbations. On the other hand, at the mesoscale, we need to consider the existence of at least two different regions, namely overdense regions and underdense regions. It is indeed well-known that matter is not evenly distributed over space, but has grouped together, forming particular patterns in which the density is much larger than on average over whole space. 
	
	To fix ideas, we consider a simplified approach assuming that those two regions are both characterized by uniform mean properties. So, according to Einstein's law of General Relativity, overdense regions present a dynamics verifying
	\begin{equation}\label{hhq1}
		G_{tt(o)} = 8\pi G T_{tt(o)}\,,
	\end{equation}
	whereas underdense regions present a dynamics verifying
	\begin{equation}\label{hhq2}
		G_{tt(u)} = 8\pi G T_{tt(u)}\,,
	\end{equation}
	where we used the subscripts $(o)$ and $(u)$ to specify in which region the tensors are evaluated (overdense and underdense, respectively). 
	
	Let us write these two equations in terms of an average and a perturbation part:
	\begin{equation}\label{hn}
		\overline{G}_{tt} + \Delta G_{tt(i)}  = 8\pi G \left(\overline{T}_{tt} + \Delta T_{tt(i)}\right)\,,
	\end{equation}
	where $(i)$ stands for $(o)$ or $(u)$. Here, $\overline{G}_{tt}$ is the first diagonal component of the Einstein tensor corresponding to the as fitted FLRW metric, and is constant over whole space. On the other hand, $\Delta G_{tt(i)}$ is the difference between the local value and the fitted one, and may differ between overdense and underdense regions. In the right hand side, $\overline{T}_{tt}$ corresponds to the average of the density over whole space (i.e., $\rho$ as defined in Eq.\ $(\ref{ro})$), while $\Delta T_{tt(i)}$ is the difference between the local density and the average one. On average over whole space, this difference cancels, but it can be significant in the two regions. Indeed, in underdense regions, the density is vanishing, meaning that with respect to the average density over whole space, the perturbation part should be of the same order of magnitude: $\left|\Delta T_{tt(u)}\right| \simeq \rho$. In overdense regions, the magnitude of the perturbation part can even be much larger, especially in later times. This is because due to the gravitational attraction, matter has grouped together in such small volumes with respect to overall space that the density there has reached values which can be larger than twice the average one (see for example \cite{Cautun}).
	
	Now, for $\Delta G_{tt(i)}$, the magnitude of this perturbation term in one region or in the other will strongly depend on how the FLRW metric (and its related Einstein tensor) has been fitted. Indeed, reminding the simple case of the earth, the height of the mountains and the depth of the valleys with respect to the sphere supposed to represent the earth depend on how this sphere has been fitted on the earth. The same thing holds for the universe and its FLRW metric. We understand however that lowering the perturbation term in one region will be at the expense of an increase of the perturbation term in the other region, and vice versa. To fix ideas, we consider here the case for which the fitting is such that the average Einstein equation reduces to the Einstein equation itself, meaning thus that the second term in the left hand side of Eq.\ $(\ref{qss})$ cancels. This thus means that the perturbation term $\Delta G_{tt(i)}$ cancels on average over whole space, and so the average Einstein equation described by Eq.\ $(\ref{qss})$ reduces to the Friedmann equation:
	\begin{equation}
		3\frac{\dot{a}^2}{a^2} = 8\pi G\rho\,.
	\end{equation}
	Combining this together with Eq.\ $(\ref{hn})$, we obtain
	\begin{equation}
		\Delta G_{tt(i)} = 8\pi G \Delta T_{tt(i)}\,.
	\end{equation}
	But at our epoch, since the order of magnitude of $\Delta T_{tt(i)}$ is that of the average density, and since this average density is related to $\overline{G}_{tt}$ through the Einstein equation of General relativity, we deduce from the last equation that the order of magnitude of $\Delta G_{tt(i)}$ is the one of $\overline{G}_{tt}$ also. In other words, perturbations are similar in order of magnitude to the average value. So, the dynamics of the expansion of the universe, which is directly described by $\overline{G}_{tt} = 3\dot{a}^2/a^2$, corresponds to an average of values that are quite different in overdense and underdense regions. The significant perturbations in the Einstein tensor field are clearly indicative of a wide variety of possible fittings in the dynamics of the universe. And as we will see in the next sections, the root cause of the apparent accelerated expansion of the universe exactly corresponds to a fitting process that significantly deviates from the one in which the cosmological constant vanishes.
	
	For completeness, let us stress that we cannot deduce directly the perturbations in the metric itself, because $\Delta G_{tt(i)}$ is related not only to those terms, but also to their first and second derivatives. We can however notice that even if at some time the metric is quite smooth over space, the dynamics between both regions has become so different over time that we may expect that the local metrics will progressively deviate from each other.
	
	
	\section{The fitting of the SNIa and the BAO probes}\label{S1}

	We investigate now the specific fitting process related to the SNIa and BAO probes. Those probes imply a fitting process being probably the most complete and direct one. It is the most complete because we really track the scale factor at different times, so we have a detailed knowledge of its evolution over time. It is also the most direct one because we determine the evolution of the scale factor by a direct measurement of this parameter itself. This differs with the CMB probe, which uses indirect measurements, without any tracking over time. This last probe therefore requires some additional assumptions, see the next section.
	
	To establish the fitting process of the SNIa and BAO probes, we need to determine the second term of the left hand side of Eq.\ $(\ref{qss})$. The analysis performed in \cite{Deledicque} will therefore be very useful. It has indeed been shown in this article that if SNIa were evenly distributed over space, then this term would cancel, and the fitting process would correspond to the expected one, described by the Friedmann equation. In reality however, SNIa are not evenly distributed over space, but occur mainly in overdense regions, in which most of the matter is concentrated. The SNIa probe is thus biased, because it does not take into account results that would be collected in underdense regions, if measurements could be performed there. This bias then implies that the second term of Eq.\ $(\ref{qss})$ does not cancel. In fact, by using the simple dichotomous modelling depicted above, admitting that overdense and underdense subregions are characterized by uniform mean properties, it has been shown by \cite{Deledicque} that it is in overdense regions that the average of $\Delta G_{tt}$ cancels, instead of in whole space:
	\begin{equation}\label{qf}
		\int_{V_o}\Delta G_{tt}\sqrt{|g_{ij}|}dV = 0\,,
	\end{equation}
	where $V = V_o + V_u$, $V_o$ being the volume of overdense regions and $V_u$ being the volume of underdense regions. This can be understood as follows: measurement results of those probes are interpreted by assuming that at the points were they are performed, the universe is described by the FLRW metric. So, since measurements are only performed in overdense regions, it is solely in such regions that this assumption is made, meaning thus that on average in these regions, $\overline{g}_{\mu\nu(o)}$ and all derived tensors (in particular $\overline{G}_{\mu\nu(o)}$) vanish. We refer to \cite{Deledicque} for a formal demonstration of this result. In other words, the FLRW metric is fitted on the overdense regions instead of over whole space. Comparing with the case of the earth, this is as if we were fitting the mean radius by performing measurements on the top of the mountains only.	
		
	The bias implies thus a completely different fitting of the FLRW metric than the one we would expect if measurements were evenly distributed trough space. Indeed, given Eq.\ $(\ref{qf})$, Eq.\ $(\ref{qss})$ reduces to
	\begin{equation}\label{qs}
		3\frac{\dot{a}^2}{a^2} + \frac{\int_{V_u}\Delta G_{tt}\sqrt{|g_{ij}|}dV}{\int_V \sqrt{|g_{ij}|}dV} = 8\pi G\rho\,.
	\end{equation}
	With respect to the usual Friedmann equation (with $\Lambda = 0$), there remains an additional term. It was shown by \cite{Deledicque} that this term can account for an apparent dark energy effect, quantitatively as well as qualitatively. In particular, it has been shown that this term can be related to a density whose order of magnitude corresponds to the one of the measured dark energy density. It has also been shown that this term can be related to a negative pressure, equal in magnitude to the value of the dark energy density.
	
	In fact, we can find back the results of \cite{Deledicque} by following the approach developed above. But instead of performing the integrations over whole space $V$, we perform them over the overdense volume $V_o$ only. In that case, Eq.\ $(\ref{az})$ would become
	\begin{equation}
		\overline{G}_{tt} + \frac{\int_{V_o}\Delta G_{tt}\sqrt{|g_{ij}|}dV}{\int_{V_o} \sqrt{|g_{ij}|}dV} = 8\pi G\frac{\int_{V_o}  T_{tt}\sqrt{|g_{ij}|}dV}{\int_{V_o} \sqrt{|g_{ij}|}dV}\,.
	\end{equation}
	But due to the bias, the second term of the left hand side cancels. Also, the right hand side corresponds to the average density in the overdense regions $\rho_o$. We thus have
	\begin{equation}\label{dd}
		\overline{G}_{tt}  = 8\pi G\rho_o = 8\pi G \left(\rho + \Delta\rho_o\right)\,,
	\end{equation}
	where $\Delta\rho_o$ represents the difference between the average density in overdense regions and the average density over whole space. It is now the term $\rho_0$ that accounts for an apparent dark energy effect. In earlier times, when heterogeneities were smaller than currently and when $\rho$ was large, this term was negligible. But over time, heterogeneities have developed, $\rho$ has decreased, and $\Delta\rho_o$ has become more important. This term is thus responsible for a deviation in the dynamics with respect to the one expected from the Friedmann equation. We refer to \cite{Deledicque} for a more in-depth analysis of this term.
	
	Moreover, it has been shown in \cite{Deledicque} that the bias that exists for the SNIa probe also exists for the BAO probe. The BAO probe leads hence to an identical fitting process, related to the same average Einstein equation, explaining why results obtained from both probes are similar.

	Let us finally notice the following about the SNIa and BAO probes. As said in the introduction of this section, these probes are complete in the sense that they track the scale factor over time. In theory, we could thus at each time deduce an average scale factor, and then link the values at different times to establish its temporal evolution. In the general case, such temporal evolution could be anything. However, it appears that the related fitting leads to a temporal evolution that really looks like the one predicted by the Friedmann equation, taking into account the existence of a non-zero cosmological constant. Therefore, in practice, the temporal evolution of the scale factor is not fitted by an unspecified curve, but on the contrary by a well-defined function, verifying the Friedmann equation, and parametrized by the cosmological constant. As shown in \cite{Deledicque}, this approach is expected to be adequate for large values of the cosmic time, but for more earlier times, some deviation could be observed, since the second term of Eq.\ $(\ref{qss})$ only tends to present a dark energy behaviour over time. Fortunately, at earlier terms, the relative importance of this second term with respect to the first one vanishes, so the deviations should be small.
	

	\section{The fitting of the CMB probe}\label{S2}

	We now investigate the fitting process related to the CMB probe. As we will see, this fitting process presents on a conceptual point of view some differences with the one related to the SNIa and BAO probes. However, despite these differences, it leads to a very similar fitted manifold, even if a slight deviation is observed. To understand this intriguing situation, it is extremely important to have a clear view on the related fitting process.
	
	Let us first briefly remind the principle of the CMB probe. The cosmic microwave background, which fills all space, is the remnant  electromagnetic radiation from an early stage of the universe. This cosmic microwave background is almost isotropic. Though, it presents slight anisotropies. When analysing those anisotropies in terms of their power spectrum, specific peaks appear. Such peaks are the signatures of some physical phenomena, which can be related to different parameters of interest, in particular the curvature of the universe, the baryon density, and the dark-matter density. After having determined these parameters from the CMB anisotropies, the dark energy density can indirectly be deduced.
	
	We first notice that the CMB probe is an indirect probe. It does not directly measure the scale factor as does the SNIa probe, but on the contrary determines it indirectly on the basis of other types of measurements. Conceptually, this is not an issue. If the physics is well-understood and well-described, the equations of the model predicting the CMB anisotropies should form a coherent set, and the knowledge of some parameters should allow us to determine some other ones.	
	
	More importantly, unlike the SNIa and BOA probes, the CMB probe does not track the scale factor at different times. This is a disadvantage. Indeed, this means that we have no information allowing to confirm (or not) that the temporal evolution of the scale factor indeed corresponds to the one predicted by the Friedmann equation. On the contrary, the CMB probe postulates that this temporal evolution will correspond to the one predicted by the Friedmann equation. As a consequence, the idealized manifold that we fit on the real universe is constrained by this postulate: it corresponds to a well-defined function, parametrized by the cosmological constant $\Lambda$, whose value is deduced from the measurements of curvature as well as the baryon and dark matter densities. 	
	
	At last, the CMB probe as it is applied suffers from a subtle but major drawback. As we insisted on in the introduction of section $\ref{Ao}$, fitting processes rely in general on averaging operations, because of the existence of perturbations. We cannot rely on the result of one single measurement to fit a manifold, because due to those perturbations, this single data cannot in general be considered as being representative of the global behaviour. Though, in the CMB probe as it is (and can be) applied in practice, there is no such averaging operation, simply because we have at our disposal only one single result. Let us clearly explain this. We fit the idealized manifold by constraining it to be related to the Friemann equation. This manifold is parametrized by the matter and dark energy densities, so the fitting process will be accomplished as these parameters are fixed. The CMB probe provides one value for each of these parameters. But the CMB measurements are most probably affected by the perturbations of the metric existing at the locations were they have been carried out, i.e., in the vicinity of the earth. These single values for the matter and dark energy densities are thus not reliable to deduce a global behaviour, this is as we were fitting the sphere representing the earth by performing just a measurement of its radius at the Everest. In some sense, we fit the idealized manifold by constraining it to exactly pass trough our particular point, which unfortunately cannot be considered as globally representative. So we should repeat the CMB measurements at other locations in space to get an average value for $\Lambda$ and the matter density.
	
	Let us justify why the perturbations of the metric in the vicinity of the earth could affect so much the fitting process. To interpret the anisotropies measured in the cosmic microwave background and deduce the different parameters, a theoretical model has been built, based on the assumption that space is globally homogeneous and isotropic, and hence that the universe's metric tensor corresponds to the FLRW tensor everywhere. It is only with such an assumption that quite simple theoretical relations could be obtained in the model. But these relations are only valid at a global scale, meaning on average if measurements were repeated at different locations. At a very specific location, these relations are not valid, because the local perturbation to this metric tensor have to be taken into account. The significance of such perturbations has already been shown above, but let us once again illustrate this by a simple example. In the Friedmann equation (assuming $\Lambda = 0$), the Hubble constant can be determined as
	\begin{equation}
		3H_0^2 = 8\pi G\rho\,.
	\end{equation}
	But if the Hubble constant is measured from the earth (being located in an overdense region), we should in fact use Eq.\ $(\ref{hhq1})$:
	\begin{eqnarray}
		3H_0^2 + \Delta G_{tt(o)} = 8\pi G \left(\rho + \Delta \rho_{o}\right)\,.
	\end{eqnarray}
	Obviously, both equations are theoretically equivalent to deduce $H_0$, because $\Delta G_{tt(o)} = 8\pi G\Delta \rho_{o}$. But when performing the measurements (on events quite close to us), if in their interpretation we neglect the perturbations of the metric tensor (and of the related derived tensors, in particular the perturbation in the Einstein tensor), what we would measure should correspond to
	\begin{eqnarray}
		3H_0^2 = 8\pi G \left(\rho + \Delta \rho_{o}\right)\,.
	\end{eqnarray}
	The second term in the right hand side does not cancel anymore with the second term of the left hand side. And as justified above, this term can be significant, it can be of the order of magnitude of what we would like to measure. Neglecting the perturbations in the metric can thus strongly affect the result.
	
	So in practice, the CMB probe suffers from the same bias as the SNIa and BOA probes do: by neglecting the perturbations of the metric tensor in the vicinity of the earth, the Einstein equation of General relativity reduces to Eq.\ $(\ref{dd})$, making appear an apparent dark energy content. Therefore, the fitting process related to the CMB probe leads to a similar result.

	The proposed explanation also provides a clear understanding of the Hubble tension. In our simplified model of above, such a Hubble tension should in fact not exist. Since the apparent dark energy is related to the average density in overdense regions (supposed to be constant through all these regions), the SNIA, the BAO as well as the CMB probes should evidence an identical value for it. But obviously, the model considered above is very simplified, in reality ovederdense regions exhibit densities that may vary from point to point. Since the SNIa and BOA probes are applied on a large set of events (somewhat distributed through the overdense regions), they truly evidence, on average, an apparent dark energy which is directly related to the average density in overdense regions. On the other hand, the CMB probe is applied only at our specific location, so the apparent dark energy that will be determined will be related to the local value of the density, and not the average through the overdense regions. As a matter of fact, there is no reason that the local density in our vicinity exactly corresponds to the average density through overdense regions. As a consequence, a different apparent dark energy should be expected. In some sense, as it is applied, the CMB probe indirectly provides a measure of our local density. Unfortunately, a direct quantitative comparison with measurements of the local density is difficult. Since in reality matter is not distributed as a continuous fluid, it is unclear how we should define our local density. Its value strongly depends on which volume it is evaluated. For very small volumes around our galaxy, its value is significantly above the critical density, but when increasing the volume, at some point it becomes smaller than the critical density (see \cite{Karachentsev}). So if this model proposes a simple theoretical explanation for the Hubble tension, a quantitative confirmation will require a different approach. This is however beyond the scope of this article.
	
	
	\section{Conclusion}
	
	Understanding how cosmic probes fit the idealized FLRW metric onto the real universe is fundamental, because it is those fittings that are analysed in practice, and any property derived from them is supposed to be applicable to the real universe as well. It is hence required to have fitting processes leading to idealized approximations being indeed representative of the real universe. In this article we have investigated the fitting processes implicitly defined by the SNIa, BAO and CMB cosmic probes. We first insisted on the fact that different fitting processes could lead to different dynamics. Therefore, we derived an averaged Einstein equation of General Relativity which, in addition to being covariant, contains an additional term depending on how the FLRW metric has been fitted. For each cosmic probe, we have then determined this additional term, and shown that it was able to account for an apparent dark energy effect. This justifies that there is no need to assume a non-zero cosmological constant to explain the observed accelerated expansion of the universe. Furthermore, an in depth understanding of the conceptual differences between the CMB probe and the two other ones allowed us to provide an explanation for the so-called Hubble tension. In summary, the investigation supports the idea that if on a global scale, the universe can be considered as being homogeneous and isotropic, local inhomogeneities have to be considered in the interpretation of the results provided by the various cosmic probes.
	


	
\end{document}